\documentclass[column,showpacs,amsmath,amssymb,amsfonts,nofootinbib,plb,]{revtex4}
\usepackage{graphicx}
\usepackage{dcolumn}
\usepackage{bm}
\usepackage{ulem}
\usepackage{overpic}
\usepackage{amssymb}
\usepackage{amsmath}
\usepackage{datetime}

\newcommand{\nn}{\nonumber}

\def\beq{\begin{equation}}
\def\eeq{\end{equation}}

\usepackage{soul}
\usepackage{cancel}

\allowdisplaybreaks[1]
\usepackage[usenames]{color}

\begin{document}

\title{JIMWLK evolution and small-${x}$ asymptotics  of $2n$-tuple Wilson line correlators}
\author{Khatiza Banu,  Mariyah Siddiqah and Raktim Abir}   
\affiliation{Department of Physics, Aligarh Muslim University, Aligarh - $202002$, India.}  
\begin{abstract}
JIMWLK equation tells how gauge invariant higher order Wilson line correlators would evolve at high energy. 
In this article we present a convenient integro-differential form of this equation, for $2n$-tuple correlator, where all real and virtual terms are explicit. 
The `real' terms correspond to splitting (say at position $z$) of this $2n$-tuple correlator to various pairs of $2m$-tuple and $(2n+2-2m)$-tuple correlators whereas `virtual' terms correspond to splitting into pairs of $2m$-tuple and $(2n-2m)$-tuple correlators. 
Kernels of virtual terms with $m=0$ (no splitting) and of real terms with $m=1$ (splitting with atleast one dipole) have poles and when integrated over $z$ they do generate ultraviolet logarithmic divergences, separately for real and virtual terms. 
Except these two cases in all other terms the corresponding  kernels, separately for real and virtual terms,  have rather soften ultraviolet singularity and when integrated over $z$ do not generate ultraviolet logarithmic divergences. 
We went on to study the solution of the JIMWLK equation for the $2n$-tuple Wilson line correlator in the strong scattering regime where all transverse distances are much larger than inverse saturation momentum and shown that it also exhibits geometric scaling like color dipole deep inside saturation region. 
\end{abstract}

 \pacs{12.38.−t}

 \date{\today ~~\currenttime}

 \maketitle
 
 \section{Introduction}
High energy scattering  in QCD \cite{Gribov:1984tu} can be most conveniently addressed using the color dipole degrees of freedom by Mueller \cite{Mueller:1994jq,Mueller:1993rr}. 
In the study of high energy scattering of a projectile parton  and a target  nucleus  the small-$x$ evolution can be introduced either in the wave function of the projectile (the parton) or in the wave function of the target (the nucleus). 
The Balitsky-Kovchegov (BK) evolution equation / Balitsky hierarchy \cite{Balitsky:1995ub,Kovchegov:1999yj} accomplishes the first while the other equivalent approach is realised by Jalilian-Marian - Iancu - McLerran - Weigert - Leonidov - Kovner (JIMWLK) \cite{JalilianMarian:1997jx, JalilianMarian:1997gr, Iancu:2001ad} evolution equation. 
In this context when estimating color averaged expectation value of certain operator one generally is in need of an appropriate weight function for the color field. 
In the Color Glass Condensate (CGC) \cite{Iancu:2000hn, Iancu:2002xk, Iancu:2003xm,Gelis:2010nm} effective theory the spatial distribution of the color sources that produce classical color field inside the large target nucleus is taken to be Gaussian. 
The JIMWLK formalism generalizes this Gaussian weight  of classical gluon field to a rapidity-dependant weight functional ${\cal W}_{\hspace{0.04cm}Y}\left[\alpha\right]$ which no longer remains Gaussian as it evolve across the energy or rapidity.  
Unlike McLerran - Venugopalan (MV) model where the weight function is Gaussian always, here the weight function has to be determined from the JIMWLK equation itself for evaluation at certain rapidity $Y$. 
All physical measurable quantities are expressed as gauge invariant operators ${\cal O}$ build with  the color field $\alpha$ and corresponding expectation values are obtained after averaging over the stochastic color field $\alpha$: 
\begin{eqnarray}
\langle {\cal O} \rangle \equiv \int {\cal D}\alpha~{\cal O}[\alpha] ~{\cal W}_{\hspace{0.04cm}Y}\left[\alpha \right]
\end{eqnarray}
The functional differential evolution equation for ${\cal W}_{\hspace{0.04cm}Y}\left[\alpha \right]$ is the JIMWLK equation and reads, 
\begin{eqnarray}
\frac{\partial}{\partial Y}~{\cal W}_{\hspace{0.04cm}Y}[\alpha]&=& {\cal H}~{\cal W}_{\hspace{0.04cm}Y}[\alpha], 
\end{eqnarray}
where $Y\equiv \ln \left(1/x\right)$ and ${\cal H}$ is the JIMWLK Hamiltonian, 
\begin{eqnarray}
 {\cal H} &\equiv& \frac{1}{2}\int_{\textit{{xy}}}\frac{\delta}{\delta \alpha^{a}_{Y}(\textit{x})}~\chi^{ab}(\textit{{x,y}})~\frac{\delta}{\delta \alpha^{b}_{Y}({y})}~. \label{JIMWLK_Hamiltonian}
 \end{eqnarray}
The integral sign with subscript $xy$, in the Hamiltonian, denotes integration over the transverse coordinates ${{x}}$ and ${{y}}$. The kernel $\chi^{ab}(\textit{{x,y}})$ is a functional of $\alpha$ upon which it  depends through the Wilson lines $e.g.$ $\tilde{U}^{}(\textit{x})$ and $\tilde{U}^{\dagger}(\textit{{x}})$ that build with $\alpha\equiv \alpha^a T^a$ in the adjoint representation, 
\begin{eqnarray}
\chi^{ab}(\textit{{x,y}}) = \frac{1}{\pi}\int \frac{d^2z}{(2\pi)^2} ~ {\cal K} (\textit{{x,y,z}}) \left(1-{\tilde U}_{\textit{{x}}}^{\dagger}  {\tilde U}_{\textit{{z}}}\right)^{fa}
\left(1-\tilde{U}^{\dagger}_\textit{{z}}\tilde{U}_\textit{{y}}\right)^{fb}~,
 \end{eqnarray}
with the transverse kernel, 
\begin{eqnarray}
{\cal K} (\textit{{x,y,z}}) \equiv \frac{(x-z).(y-z)}{(x-z)^2(z-y)^2}~,
\end{eqnarray}
and, $e.g.$,  
\begin{eqnarray}
{\tilde U}_{\textit{{x}}}^{\dagger}  \equiv {\tilde U}^{\dagger}(\textit{{x}})= {\cal P}\exp\left(ig\int dx^- \alpha^a(x^-, x)T^a\right)~,
\end{eqnarray}
where ${\cal P}$ denotes path ordering in $x^-$, integration over $x^-$ runs over the longitudinal extent of the hadron, which increases with $Y$.   Action of functional derivative on Wilson lines in  adjoint representation reads as, 
 \begin{eqnarray}
  \frac{\delta}{\delta \alpha^{a}_\tau (y)}{\tilde U}^\dagger(\textit{{x}})=ig\delta^{(2)}(\textit{{x}}-\textit{{y}})T^a {\tilde U}^\dagger_{\textit{{x}}}
  \qquad 
\frac{\delta}{\delta \alpha^{a}_\tau (y)}{\tilde U}(\textit{{x}})=-ig\delta^{(2)}(\textit{{x}}-\textit{{y}}){\tilde U}_\textit{{x}}T^a ~. 
 \end{eqnarray}  
Within the leading logarithmic accuracy, the CGC effective theory prescribes following energy evolution for general gauge invariant operator 
${\cal O}$, 
\begin{eqnarray}
\frac{\partial}{\partial Y} \langle \hat {\cal O} \rangle_{Y} = \langle {\cal H} \hat {\cal O}  \rangle_{Y}~. 
\end{eqnarray}
%
The brackets refer to average over color fields in the target nucleus, properly accompanied by the rapidity dependent CGC weight function as mentioned before and ${\cal H}$ is the JIMWLK Hamiltonian, 
\begin{eqnarray}
{\cal H} \equiv -\frac{1}{16\pi^3}\int_{z} {\cal M}_{\textit{{xyz}}}\left(1+\tilde{U}^{\dagger}_\textit{\textbf{x}}\tilde{U}_\textit{\textbf{y}}  - \tilde{U}^{\dagger}_\textit{\textbf{x}}\tilde{U}_\textit{\textbf{z}} - \tilde{U}^{\dagger}_\textit{\textbf{z}}\tilde{U}_\textit{\textbf{y}}\right)^{ab}\frac{\delta}{\delta \alpha_{x}^{a}}\frac{\delta}{\delta \alpha_{y}^{b}} ~, \label{JIMWLK-02}
\end{eqnarray}
where ${\cal M}_{\textit{{xyz}}}$ is the dipole kernel, 
\begin{eqnarray}
{\cal M}_{\textit{{xyz}}} \equiv \frac{\left({\textit{{x}}} - {\textit{{y}}}\right)^2}{({\textit{{x}}} - {\textit{{z}}})^2({\textit{{z}}} - {\textit{{y}}})^2}~.
\end{eqnarray}
In this paper we have studied $2n$ point operator \cite{Ayala:2014nza,Shi:2017gcq}\footnote{The KLWMIJ evolution equation corresponds to the evolution of the projectile weight functional in the scattering of a dilute projectile on a dense target.   
The KLWMIJ equation is  dual to JIMWLK evolution of the same object in the scattering of a dense projectile on a dilute target \cite{Altinoluk:2014mta}. Computation of  the KLWMIJ evolution for correlators of an arbitrary number of Wilson lines have been done in \cite{Altinoluk:2014twa}.}
constructed from the Wilson lines in the fundamental representation: 
\begin{eqnarray}
{\cal S}^{(2n)}\equiv \frac{1}{N_c}{\rm Tr}\left[U(x_{1})U^{\dagger}(x_{2})U(x_{3})U^{\dagger}(x_{4}) ... U(x_{2n-1})U^{\dagger}(x_{2n})\right]~.
\end{eqnarray}
The color dipole corresponds to $n=1$ and is a quark-anti quark pair in overall color singlet 
state. The operator for color dipole contains two Wilson lines in their fundamental representation, 
\begin{eqnarray}
{\cal S}^{(2)}\equiv \frac{1}{N_c}{\rm Tr}\left[U(x_{1})U^{\dagger}(x_{2})\right]~.
\end{eqnarray}
The evolution equation for the dipole is simple, and contains only two terms
both having identical kernels, leads to Balitsky-Kovchegov equation in the large-$N_c$ limit. 
Next higher point correlators are {\it color quadrupole} \cite{Dominguez:2011gc} and {\it color sextupole} \cite{Iancu:2011ns} that contain four and six Wilson lines respectively as, 
\begin{eqnarray}
{\cal S}^{(4)}\equiv \frac{1}{N_c}{\rm Tr}\left[U(x_{1})U^{\dagger}(x_{2})U(x_{3})U^{\dagger}(x_{4})\right]~,
\end{eqnarray}
and, 
\begin{eqnarray}
{\cal S}^{(6)}\equiv \frac{1}{N_c}{\rm Tr}\left[U(x_{1})U^{\dagger}(x_{2})U(x_{3})U^{\dagger}(x_{4})U(x_{5})U^{\dagger}(x_{6})\right]~.
\end{eqnarray}
The evolution equation for both the operator have been derived in \cite{Iancu:2011ns}. 
In this article we also studied evolution for color octupole operator, 
\begin{eqnarray}
{\cal S}^{(8)}\equiv \frac{1}{N_c}{\rm Tr}\left[U(x_{1})U^{\dagger}(x_{2})U(x_{3})U^{\dagger}(x_{4})U(x_{5})U^{\dagger}(x_{6})U(x_{7})U^{\dagger}(x_{8})\right]~,
\end{eqnarray}
and color decapole, 
\begin{eqnarray}
{\cal S}^{(10)}\equiv \frac{1}{N_c}{\rm Tr}\left[U(x_{1})U^{\dagger}(x_{2})U(x_{3})U^{\dagger}(x_{4})U(x_{5})U^{\dagger}(x_{6})U(x_{7})U^{\dagger}(x_{8})U(x_{9})U^{\dagger}(x_{10})\right]~,
\end{eqnarray}
that contain eight and ten Wilson lines in their fundamental representation.
\\ 

This paper is organised as follows. In the section II we present a convenient integro-differential form of this equation, for $2n$-tuple correlator. A few special cases $e.g.$ dipole, quadrupole and sextupole amplitudes including the evolution of  octupole and decapole have been jotted down in the appendix.  In section III we have analysed the generic kernel of the equation and demonstrate that it is ultraviolet safe when the splitting does not involve any dipole. In section IV
we solve the equation in the unitarity limit and find the solution for $2n$-tuple correlator. Finally we conclude in section V.

\section{High energy evolution of color {\bf {\it $2n$}}-tuple correlator}
Here we present the explicit integro-differential   form of the JIMWLK evolution equation for $2n$-tuple Wilson line correlator. This is derived by operating the JIMWLK Hamiltonian in  Eq.\eqref{JIMWLK-02} on a general gauge invariant operator with $2n$ Wilson lines in their fundamental representation. All virtual terms are generated by the first two terms of the Hamiltonian whereas the last two terms produce the real terms. All the terms in the above equation are leading in $N_c$ while all the sub-leading terms of the order of $1/N_c^2$, generated at the intermediate steps, cancel after summing up all the contributions in the final equation
The details of derivation for octupole and decapole are in the appendix. We note that {\it whenever the transverse position index notation is greater than $2n$ it should be realised with its modulo $e.g.$ $x_{2n+k}\equiv x_{k}$}. 
\begin{eqnarray}
&&~~~~ \frac{\partial}{\partial Y} {\rm Tr}\left[U(x_{1})U^{\dagger}(x_{2})U(x_{3})U^{\dagger}(x_{4}) ... U(x_{2n-1})U^{\dagger}(x_{2n})\right]=\frac{\bar \alpha_{s}}{4\pi}\left(\frac{1}{1+\delta_{n,1}}\right) \times \nn \\
&& \int_{z} \sum_{k=0}^{\lfloor n/2 \rfloor -1} ~ \sum_{l=0}^{n-1} 
{\cal K}_{(2l+1;2l+2k+1)}^{(2l;2l+2k+2)}{\rm Tr}\left[U(x_{2l+1})U^{\dagger}(x_{2l+2}) ... U(x_{2l+1+2k})U^{\dagger}(z)\right]{\rm Tr}\left[U(z)U^{\dagger}(x_{2l+2k+2}) ... U(x_{2l-1})U^{\dagger}(x_{2l})\right]\nn \\
&&+\sum_{k=0}^{\lfloor n/2 \rfloor -1}~\sum_{l=0}^{n-1} {\cal K}_{(2l+2;2l+2k+2)}^{(2l+1;2l+2k+3)}~{\rm Tr}\left[U^{\dagger}(x_{2l+2})U(x_{2l+3}) ... U^{\dagger}(x_{2l+2+2k})U(z)\right]{\rm Tr}\left[U^{\dagger}(z)U(x_{2l+2k+3}) ... U^{\dagger}(x_{2l})U(x_{2l+1})\right]  \nn  \\
&&+\sum_{k=0}^{\lceil n/2 \rceil -2} ~ \sum_{l=0}^{n-1}{\cal K}_{(2l+1;2l+2k+2)}^{(2l;2l+2k+3)}~{\rm Tr}\left[U(x_{2l+1})U^{\dagger}
(x_{2l+2})... U^{\dagger}(x_{2l+1+2k+1})\right]{\rm Tr}
\left[U(x_{2l+2k+3})U^{\dagger}(x_{2l+2k+4}).....U^{\dagger}(x_{2l})\right]\nn\\
&&+\sum_{k=0}^{\lceil n/2 \rceil -2} ~ \sum_{l=0}^{n-1} {\cal K}_{(2l+2;2l+2k+3)}^{(2l+1;2l+2k+4)}~{\rm Tr}
\left[U^{\dagger}(x_{2l+2})U(x_{2l+3})...U(x_{2l+2k+3})\right]
{\rm Tr}\left[U^{\dagger}(x_{2l+2k+4})U(x_{2l+2k+5})...U(x_{2l+1})\right]\nn\\
&&+ \delta_{1, n\bmod{2}} ~ \sum_{l=0}^{\lceil n/2 \rceil -1} 
{\cal K}_{(2l+1;2l+n)}^{(2l;2l+n+1)}{\rm Tr}\left[U(x_{2l+1})U^{\dagger}(x_{2l+2}) ... U(x_{2l+n})U^{\dagger}(z)\right]{\rm Tr}\left[U(z)U^{\dagger}(x_{2l+n+1}) ... U(x_{2l-1})U^{\dagger}(x_{2l})\right]\nn \\
&&+\delta_{1, n\bmod{2}}~ \sum_{l=0}^{\lceil n/2 \rceil -2}  {\cal K}_{(2l+2;2l+n+1)}^{(2l+1;2l+n+2)}~{\rm Tr}\left[U^{\dagger}(x_{2l+2})U(x_{2l+3}) ... U^{\dagger}(x_{2l+n+1})U(z)\right]{\rm Tr}\left[U^{\dagger}(z)U(x_{2l+n+2}) ... U^{\dagger}(x_{2l})U(x_{2l+1})\right]  \nn  \\
&&+ \delta_{0, n\bmod{2}}\sum_{l=0}^{n/2-1}{\cal K}_{(2l+1;2l+n)}^{(2l;2l+n+1)}~{\rm Tr}\left[U(x_{2l+1})U^{\dagger}
(x_{2l+2})... U^{\dagger}(x_{2l+n})\right]{\rm Tr}
\left[U(x_{2l+n+1})U^{\dagger}(x_{2l+n+2}).....U^{\dagger}(x_{2l})\right] \nn \\
&&+ \delta_{0, n\bmod{2}}\sum_{l=0}^{n/2-1} {\cal K}_{(2l+2;2l+n+1)}^{(2l+1;2l+n+2)}~{\rm Tr}
\left[U^{\dagger}(x_{2l+2})U(x_{2l+3})...U(x_{2l+n+1})\right]
{\rm Tr}\left[U^{\dagger}(x_{2l+n+2})U(x_{2l+n+3})...U(x_{2l+1})\right]\nn\\
&&-~{\cal P}_{2n} ~{\rm Tr}\left[U(x_{1})U^{\dagger}(x_{2})U(x_{3})U^{\dagger}(x_{4}) ... U(x_{2n-1})U^{\dagger}(x_{2n})\right]~,   \label{the.equation}
 \end{eqnarray}
where the kernels are of two type: first one is, 
\begin{figure}
  \includegraphics[width=6in, angle=0]{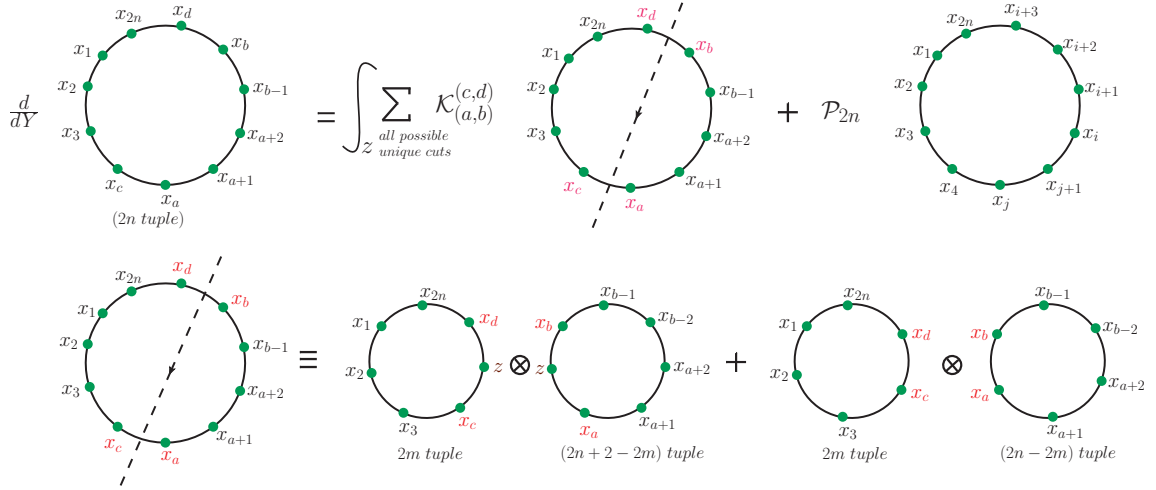}
 \caption{Schematic representation of the JIMWLK evolution for $2n$-tuple Wilson line correlator as given in Eq.\eqref{the.equation}. Circles corresponds to general multi order Wilson line correlator;  the circle with a cut line refer to terms corresponds to both real and virtual splitting while ${\cal K}$ and ${\cal P}$ are associated kernels defined in Eq.\eqref{the.karnel001} and Eq.\eqref{the.karnel002}.}
 \label{Fig01}
 \end{figure}
 \begin{eqnarray}
{\cal K}_{(a;b)}^{(c;d)}&\equiv &\frac{(x_a-x_d)^2}{(x_a-z)^2(z-x_d)^2}+\frac{(x_b-x_c)^2}{(x_b-z)^2(z-x_c)^2}-\frac{(x_a-x_b)^2}{(x_a-z)^2(z-x_b)^2}-\frac{(x_c-x_d)^2}{(x_c-z)^2(z-x_d)^2} ~,
 \label{the.karnel001} 
 \end{eqnarray}
the stand alone kernel in the last term, 
 \begin{eqnarray}
{\cal P}_{2n}&\equiv & \sum_{j=1}^{2n} \frac{(x_j-x_{j+1})^2}{(x_j-z)^2(z-x_{j+1})^2}~.
 \label{the.karnel002} 
 \end{eqnarray}
 
The terms involving  $U(z)$ or  $U^{\dagger}(z)$ are the real terms which describes splitting, at position $z$, of this $2n$-tuple correlator to pair of $2m$-tuple and $(2n+2-2m)$-tuple correlators.  These terms have been generated by the last two terms of the Hamiltonian. 
The `virtual' terms correspond to splitting into pairs of $2m$-tuple and $(2n-2m)$-tuple correlators and are generated by the first two terms of the Hamiltonian and are necessary for the probability conservation and unitarity restoration.  All the possible ultraviolet (i.e. short distance) divergences in the dipole kernel get cancelled out between the virtual and real terms, because of the probability conservation together with the property of color transparency. \\
 
The above equation suffers the problem in the sense 
 that it is not a closed equation because the right hand side includes higher-point correlations. The way to deal with this difficulty is the same as for the Balitsky-Kovchegov equation assuming that, for a large nucleus, these correlators can be factored as products of correlators involving only one trace at a time when the  large-$N_ c$ limit is taken. \\
 
 A careful look in the equation reveals that the terms in the equations are broadly of type: terms that corresponds to splitting  (either real or virtual) and the $2n$-tuple term itself. While the class of kernels ${\cal K}_{(a;b)}^{(c;d)}$ are of the former, the ${\cal P}_{2n}$ is associated with the last term $i.e.$ the $2n$-tuple term. This is schematically represented in Fig.[\ref{Fig01}]: Circles corresponds to general multi order Wilson line correlator;  the circle with a cut line refer to terms corresponds to both real and virtual splitting. As evident from figure the kernel ${\cal K}_{(a;b)}^{(c;d)}$ are defined by set of four points $(x_a,x_b,x_c,x_d)$ where the actual splitting happened for that particular term. When  a dipole is produced in a real splitting two of the four position coordinate would be identical. Next we documented the evolution equations for several special cases upto decapole $(n=5)$.

\section{Kernel for 2n-tuple correlator}
 \subsubsection{Dipole evolution}
 \noindent For dipole $n=1$,  the evolution equation becomes, 
 ~~~~~~~~~~~~~~~~
 \begin{eqnarray}
   &&\frac{\partial}{\partial Y} \langle{\rm Tr}\left[U(x_{1})U^{\dagger}(x_{2})\right]\rangle_{Y} \nn\\
&=&\frac{\bar \alpha_{s}}{4\pi} \frac{1}{2} \int_{z}{\cal K}_{(1;1)}^{(2;2)}\langle
{\rm Tr}\left[U(x_1)U^{\dagger}(z)\right]
{\rm Tr}\left[U(z)U^{\dagger}(x_2)\right]\rangle_Y -\left({\cal P}_{(1,2)}+{\cal P}_{(2,1)}\right)
\langle{\rm Tr}\left[U(x_1)U^{\dagger}(x_2)\right]\rangle_Y 
 \end{eqnarray}
Now the kernels are,  
 \begin{eqnarray}
 {\cal K}_{(1;1)}^{(2;2)}&=&\frac{(x_1-x_2)^2}{(x_1-z)^2(z-x_2)^2}+\frac{(x_1-x_2)^2}{(x_1-z)^2(z-x_2)^2}-\frac{(x_1-x_1)^2}{(x_1-z)^2(z-x_1)^2}-\frac{(x_2-x_2)^2}{(x_2-z)^2(z-x_2)^2} \nn \\
 &=&2\frac{(x_1-x_2)^2}{(x_1-z)^2(z-x_2)^2}
\end{eqnarray}
and, 
\begin{eqnarray}
{\cal P}_{(1,2)}={\cal P}_{(2,1)}= \frac{(x_1-x_2)^2}{(x_1-z)^2(z-x_2)^2}
\end{eqnarray} 
Important simplifications and factorizations
 occur in the large-$N_c$ limit that leads to the BK equation for color dipole $S(x_1,x_2)\equiv (1/N_c) \langle{\rm Tr}\left[U(x_{1})U^{\dagger}(x_{2})\right]\rangle_{Y}$, 
 \begin{eqnarray}
   &&\frac{\partial}{\partial Y} S(x_1,x_2) =\frac{\bar \alpha_{s}}{4\pi} \int_{z}  \frac{(x_1-x_2)^2}{(x_1-z)^2(z-x_2)^2} \left[S(x_1,z) S(z,x_2) - S(x_1,x_2)\right]
 \end{eqnarray}
within a mean field approximation. Now the kernel of dipole evolution equation is of the following form, 
\begin{eqnarray}
{\cal K}_{}&\equiv &\frac{(x_1-x_2)^2}{(x_1-z)^2(z-x_2)^2}~,  
\end{eqnarray}
this kernel together with the measure $d^2z$ is $SL(2,C)$ invariant. The kernel is singular at $z=x_1$ and $z=x_2$ as well. This two poles when integrated over $z$, in the strong scattering regime where all transverse distances are much larger than inverse saturation momentum, generate logarithmic divergences as, 
\begin{eqnarray}
\int_{\rho} d^2z \frac{(x_1-x_2)^2}{(x_1-z)^2(z-x_2)^2}=2\pi \ln \frac{|x_1-x_2|^2}{\rho^2}
\end{eqnarray}
where $1/\rho^2$ is some ultraviolet cut off which usually taken to be the saturation scale $Q_s^2(Y)$. In the limit $\rho\rightarrow 0$ (or $Q_s^2(Y)\rightarrow \infty$) integral is logarithmic divergent. For this particular case of dipole equation the kernels for both real and virtual terms are identical and this short distance $i.e.$ ultraviolet singularities between 
real and virtual terms would cancel each other in the overall solution. 

\subsubsection{Splitting with atleast one dipole: real terms}
\begin{figure}
  \includegraphics[width=6in, angle=0]{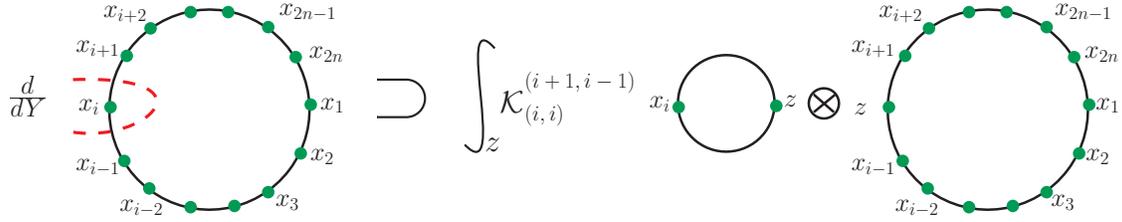}\label{}.
 \caption{Real splitting of 2n-tuple into dipole and another 2n-tuple correlator. }
 \label{}
 \end{figure}
When a $2n$-tuple splits in two lower order color multi-poles of which one 
is a dipole then also the kernel and the integral over the kernel, in the strong scattering regime, both are ultraviolet divergent. It is evident from Eq.\eqref{the.karnel001}  and Eq.\eqref{the.karnel002} that  the kernels for a general real spiting term where atleast one daughter is a dipole (e.g., $\langle{\rm Tr}\left[U(x_{i})U^{\dagger}(z)\right]$), can be written as (note for this particular case $x_{a}=x_{b}=x_{i}$), 
\begin{eqnarray}
{\cal K}_{(x_i;x_i)}^{(x_{i+1};x_{i-1})}(z)&\equiv &\frac{(x_i-x_{i-1})^2}{(x_i-z)^2(z-x_{i-1})^2}+\frac{(x_i-x_{i+1})^2}{(x_i-z)^2(z-x_{i+1})^2}-\frac{(x_i-x_i)^2}{(x_i-z)^2(z-x_i)^2}
-\frac{(x_{i+1}-x_{i-1})^2}{(x_{i+1}-z)^2(z-x_{i-1})^2}~, \nn \\
&=&\frac{(x_i-x_{i-1})^2}{(x_i-z)^2(z-x_{i-1})^2}+\frac{(x_i-x_{i+1})^2}{(x_i-z)^2(z-x_{i+1})^2}-\frac{(x_{i+1}-x_{i-1})^2}{(x_{i+1}-z)^2(z-x_{i-1})^2}~,
\end{eqnarray}
when integrated over $z$, in the strong scattering regime where all transverse distances are much larger than inverse saturation momentum, this kernel again generates logarithmic divergences as, 
\begin{eqnarray}
\int_{\rho} d^2z ~ {\cal K}_{(i;i)}^{(x_{i+1};x_{i-1})} &=& \int d^2z  \frac{(x_i-x_{i-1})^2}{(x_i-z)^2(z-x_{i-1})^2}+\frac{(x_i-x_{i+1})^2}{(x_i-z)^2(z-x_{i+1})^2}-\frac{(x_{i+1}-x_{i-1})^2}{(x_{i+1}-z)^2(z-x_{i-1})^2}~, \nn  \\
&=& 2\pi \ln \frac{|x_i-x_{i-1}|^2}{\rho^2} + 2\pi \ln \frac{|x_i-x_{i+1}|^2}{\rho^2}-2\pi \ln \frac{|x_{i+1}-x_{i-1}|^2}{\rho^2} ~, \nn \\
&=& 2\pi \ln \frac{|x_i-x_{i-1}|^2 |x_i-x_{i+1}|^2}{|x_{i+1}-x_{i-1}|^2 \rho^2}~. \label{nologdiv0001}
\end{eqnarray}
Clearly, in the limit $\rho \rightarrow 0$ the Eq.\eqref{nologdiv0001} is divergent. This is also true 
for the virtual term for which the $2n$-tuple correlator is not splitting into daughters $i.e.$ (the last term in Eq.\eqref{the.equation}),   
\begin{eqnarray}
 \int_{\rho} d^2z ~ {\cal P}_{(j;j+1)}= 2\pi \ln \prod_{1}^{2n} \frac{|x_j-x_{j+1}|^2}{\rho^2}~.
\end{eqnarray} 
This particular term is generated from the  first term (identity) and second term of the Hamiltonian in Eq.\eqref{JIMWLK-02}, and also clearly divergent in the ultraviolet limit $i.e.$ in the limit $\rho \rightarrow 0$.
 
\subsubsection{Real splitting without a daughter dipole}
As shown in Fig.[\ref{Fig01}], $(x_a,x_b)$ and $(x_c,x_d)$ are the pair of transverse position through which the splitting occurs either for real terms or for virtual terms. 
Interestingly when the real splitting do not involve any dipole or its not a virtual splitting  of two daughter of identical order ($2n$-tuple splits into two $n$-tuples), generally the kernel  would not show up any logarithmic divergence after the integration due to  cancellation between terms, 
\begin{eqnarray}
\int_{\rho} d^2z ~ {\cal K}_{(a;b)}^{(c;d)} &=& \int d^2z  \frac{(x_a-x_d)^2}{(x_a-z)^2(z-x_d)^2}+\frac{(x_b-x_c)^2}{(x_b-z)^2(z-x_c)^2}-\frac{(x_a-x_b)^2}{(x_a-z)^2(z-x_b)^2}
-\frac{(x_c-x_d)^2}{(x_c-z)^2(z-x_d)^2} ~, \nn  \\
&=& 2\pi \ln \frac{|x_a-x_d|^2}{\rho^2} + 2\pi \ln \frac{|x_b-x_c|^2}{\rho^2}-2\pi \ln \frac{|x_a-x_b|^2}{\rho^2}- 2\pi \ln \frac{|x_c-x_d|^2}{\rho^2}~, \nn \\
&=& 2\pi \ln \frac{|x_a-x_d|^2 |x_b-x_c|^2}{|x_a-x_b|^2 |x_c-x_d|^2} ~.\label{nologdiv76633}
\end{eqnarray}
Eq.\eqref{nologdiv76633} is independent of the $UV$ cut off $\rho$ and hence ultraviolet finite. 

\section{2n-tuple correlator in the unitary limit}
In the strong regime one may drop all the real terms in 
Eq.\eqref{the.equation} because they are order higher than their counter virtual terms. The equation can be written as, 
 \begin{eqnarray}
&&~~~~ \frac{\partial}{\partial Y} {\cal S}\left(x_{1}, x_{2}, x_{3}, x_{4} ... x_{2n-1}, x_{2n}\right)=\frac{\bar \alpha_{s}}{4\pi}\left(\frac{1}{1+\delta_{n,1}}\right) \times \nn \\
&&+\sum_{k=0}^{\lceil n/2 \rceil -2} ~ \sum_{l=0}^{n-1}{\cal Q}_{(2l+1;2l+2k+2)}^{(2l;2l+2k+3)}~{\cal S}^{(2k+2)}\left(x_{2l+1},x_{2l+2} ... x_{2l+1+2k+1}\right){\cal S}^{(2n-2k-2)}
\left(x_{2l+2k+3}, x_{2l+2k+4}... x_{2l}\right)\nn\\
&&+\sum_{k=0}^{\lceil n/2 \rceil -2} ~ \sum_{l=0}^{n-1} {\cal Q}_{(2l+2;2l+2k+3)}^{(2l+1;2l+2k+4)}~{\cal S}^{(2k+2)}
\left(x_{2l+2}, x_{2l+3} ... x_{2l+2k+3}\right)
{\cal S}^{(2n-2k-2)}\left(x_{2l+2k+4},x_{2l+2k+5}... x_{2l+1}\right)\nn\\
&&+ \delta_{0, n\bmod{2}}\sum_{l=0}^{n/2-1}{\cal Q}_{(2l+1;2l+n)}^{(2l;2l+n+1)}~{\cal S}^{(n)}\left(x_{2l+1}, 
x_{2l+2} ... x_{2l+n}\right){\cal S}^{(n)}
\left(x_{2l+n+1},x_{2l+n+2} ...  x_{2l}\right) \nn \\
&&+ \delta_{0, n\bmod{2}}\sum_{l=0}^{n/2-1} {\cal Q}_{(2l+2;2l+n+1)}^{(2l+1;2l+n+2)}~{\cal S}^{(n)}
\left(x_{2l+2}, x_{2l+3}...x_{2l+n+1}\right)
{\cal S}^{(n)}\left(x_{2l+n+2},x_{2l+n+3}...x_{2l+1}\right)\nn\\
&&~~~~~~~~~~~-{\cal R}~{\cal S}^{(2n)}\left(x_{1}, x_{2}, x_{3}, x_{4} ... x_{2n-1}, x_{2n}\right) \nn \\  \label{the.equation145}
 \end{eqnarray}
where ${\cal Q}$'s are defined as, 
\begin{eqnarray}
{\cal Q}_{a;b}^{c;d}\equiv \int_{\rho} d^2z ~ {\cal K}_{(a;b)}^{(c;d)} &=&  2\pi \ln \frac{|x_a-x_d|^2 |x_b-x_c|^2}{|x_a-x_b|^2 |x_c-x_d|^2} \label{nologdiv}
\end{eqnarray}
and do not explicitly depends on any infra-red cut whereas the factor ${\cal R}$ defined as, 
\begin{eqnarray}
{\cal R}\equiv \int_{\rho} d^2z ~ {\cal P}_{(j;j+1)}= 2\pi \ln \prod_{1}^{2n} \frac{|x_j-x_{j+1}|^2}{\rho^2}
\end{eqnarray} 
 Eq. \eqref{the.equation145} can further be simplified to, 
\begin{eqnarray}
&&~~~~ \frac{\partial}{\partial Y} {\cal S}\left(x_{1}, x_{2}, x_{3}, x_{4} ... x_{2n-1}, x_{2n}\right)=\frac{\bar \alpha_{s}}{4\pi} \left(\frac{1}{1+\delta_{n,1}}\right) \times \nn \\
&&+\sum_{k=0}^{\lceil n/2 \rceil -1} ~ \sum_{l=1}^{2n}2\pi \ln \frac{|x_{l}-x_{l+2k}|^2 |x_{l+2k-1}-x_{l-1}|^2}{|x_{l}-x_{l+2k-1}|^2 |x_{l-1}-x_{l+2k}|^2} ~{\cal S}^{(2k)}\left(x_{l},x_{2l+2} ... x_{l+2k-1}\right){\cal S}^{(2n-2k)}
\left(x_{l+2k}, x_{l+2k+1}... x_{x_{l-1}}\right)\nn\\
&&+ \delta_{0, n\bmod{2}}\sum_{l=1}^{n}2\pi \ln \frac{|x_{l}-x_{l+n}|^2 |x_{l+n-1}-x_{l-1}|^2}{|x_{l}-x_{l+n-1}|^2 |x_{l-1}-x_{l+n}|^2}~{\cal S}^{(n)}\left(x_{l}, 
x_{l+1} ... x_{l+n-1}\right){\cal S}^{(n)}
\left(x_{l+n},x_{l+n+2} ...  x_{l-1}\right) \nn \\
&&~~~~~~~~~~~-2\pi~{\cal S}^{(2n)}\left(x_{1}, x_{2}, x_{3}, x_{4} ... x_{2n-1}, x_{2n}\right) \ln \prod_{j=1}^{2n}  |x_j-x_{j+1}|^2 Q_s^2(Y) \nn \\  \label{the.equation1}
 \end{eqnarray} 
where in the last term we take the cutoff $\rho$ to be the inverse saturation momenta $Q_s(Y)$ that  explicitly depends on the rapidity. 
In the limit $Y\rightarrow \infty$ the last term would only survive, 
\begin{eqnarray}
&&~~~~ \frac{\partial}{\partial Y} \ln {\cal S}\left(x_{1}, x_{2}, x_{3}, x_{4} ... x_{2n-1}, x_{2n}\right)=-\frac{\bar \alpha_{s}}{2}~\left(\frac{1}{1+\delta_{n,1}}\right)~ \ln \prod_{j=1}^{2n}  |x_j-x_{j+1}|^2 Q_s^2(Y) \nn \\  \label{the.equation1}
 \end{eqnarray} 
this equation can be solve to get the Levin-Tuchin asymptotic solution for $2n$-tuple  Wilson line correlator in the unitarity limit,  
\begin{eqnarray}
{\cal S}\left(x_{1}, x_{2}, x_{3}, x_{4} ... x_{2n-1}, x_{2n}\right)
= S^{(2n)}_{0}\exp\left[-\frac{1+2i\nu_0}{2(1+\delta_{n,1})\chi(0,\nu_0)} \ln^2 \left(\prod_{j=1}^{2n}  |x_j-x_{j+1}|^2 Q_s^2(Y) \right)\right]~. 
 \end{eqnarray}

\section{Conclusion} 
In this article we present a convenient, explicit and integro-differential  form of the JIMWLK equation for general $2n$-tuple correlator in their fundamental representation. 
The `real' terms, that correspond to splitting of this $2n$-tuple correlator to various pairs of $2m$-tuple and $(2n+2-2m)$-tuple correlators, and the `virtual' terms, that correspond to splitting into pairs of $2m$-tuple and $(2n-2m)$-tuple correlators, are explicit in this integro-differential equation. 
We also presented the evolution equations, for several special cases: dipole $(n=1)$, quadrupole $(n=2)$, sextuple $(n=3)$, octupole $(n=4)$ and finally for decapole $(n=5)$. 
In this paper we have also shown that except the two cases: $m=0$ and $m=1$ for real and virtual terms respectively, where the splitting involves atleast a dipole or the term that is having the $2n$-tuple correlator itself (the last term of the Eq.\eqref{the.equation}), in all other cases the corresponding  kernels, separately for real and virtual terms,  have no ultraviolet singularity and therefore do not generate ultraviolet logarithmic divergences. 
We also study the solution of the general evolution equation in the unitarity limit $i.e.$ the strong scattering regime where all transverse distances are much larger than inverse saturation momentum.  
The solution exhibits complete geometric scaling similar to  color dipole in deep inside saturation region. 

 \begin{acknowledgements}
This work was supported in part by the University Grants Commission under UGC-BRS Research Start-Up-Grant grant number F.$30$-$310/2016$~(BSR). 
 \end{acknowledgements}
 
 \section*{Appendix}
 \subsection{Quadrupole}
For quadrupole we get $n=2$,
\begin{eqnarray}
 &&\frac{\partial}{\partial Y} \langle{\rm Tr}\left[U(x_{1})U^{\dagger}(x_{2})
U(x_{3})U^{\dagger}(x_{4}) \right]\rangle_{Y} \nn\\
&=&\frac{\bar \alpha_{s}}{4\pi}\int_{z}{\cal K}_{(1;1)}^{(4;2)}\langle
{\rm Tr}\left[U(x_1)U^{\dagger}(z)\right]
{\rm Tr}\left[U(z)U^{\dagger}(x_2)U(x_3)U^{\dagger}(x_4)\right]\rangle_Y\nn\\
&&+{\cal K}_{(2;2)}^{(1;3)}\langle
{\rm Tr}\left[U^{\dagger}(x_2)U(z)\right]
{\rm Tr}\left[U^{\dagger}(z)U(x_3)U^{\dagger}(x_4)U(x_1)\right]\rangle_Y\nn\\
&&+{\cal K}_{(3;3)}^{(2;4)}\langle
{\rm Tr}\left[U(x_3)U^{\dagger}(z)\right]
{\rm Tr}\left[U(z)U^{\dagger}(x_4)U(x_1)U^{\dagger}(x_2)\right]\rangle_Y\nn\\
&&+{\cal K}_{(4;4)}^{(3;1)}\langle
{\rm Tr}\left[U^{\dagger}(x_4)U(z)\right]
{\rm Tr}\left[U^{\dagger}(z)U(x_1)U^{\dagger}(x_2)U(x_3)\right]\rangle_Y\nn\\
&&+{\cal K}_{(1;2)}^{(4;3)}\langle{\rm Tr}\left[U(x_1)U^{\dagger}(x_2)\right]
{\rm Tr}\left[U(x_3)
U^{\dagger}(x_4)\right]\nn\\
&&+{\cal K}_{(3;2)}^{(4;1)}\langle{\rm Tr}\left[U(x_3)U^{\dagger}(x_2)\right]
{\rm Tr}\left[U(x_1)
U^{\dagger}(x_4)\right]\nn\\
&&-{\cal P}_{4}\langle{\rm Tr}\left[U(x_1)U^{\dagger}(x_2)U(x_3)U^{\dagger}(x_4)\right]
\end{eqnarray}
~~~~~~~~~~~~~~~~~~~~
\subsection{Sextupole}
For sextupole $n=3$, 
\begin{eqnarray}
 &&\frac{\partial}{\partial Y} {\rm Tr}\left[U(x_{1})U^{\dagger}(x_{2})
U(x_{3})U^{\dagger}(x_{4}) U(x_{5})U^{\dagger}(x_{6})\right] \nn\\
&=&\frac{\bar \alpha_{s}}{4\pi}\int_{z}{\cal K}_{(1;1)}^{(6;2)}\langle
{\rm Tr}\left[U(x_1)U^{\dagger}(z)\right]
{\rm Tr}\left[U(z)U^{\dagger}(x_2)U(x_3)U^{\dagger}(x_4)U(x_5)U^{\dagger}(x_6)
\right]\rangle_Y\nn\\
&&+{\cal K}_{(2;2)}^{(1;3)}\langle
{\rm Tr}\left[U^{\dagger}(x_2)U(z)\right]
{\rm Tr}\left[U^{\dagger}(z)U(x_3)U^{\dagger}(x_4)U(x_5)U^{\dagger}(x_6)U(x_1)
\right]\rangle_Y\nn\\
&&+{\cal K}_{(3;3)}^{(2;4)}\langle
{\rm Tr}\left[U(x_3)U^{\dagger}(z)\right]
{\rm Tr}\left[U(z)U^{\dagger}(x_4)U(x_5)U^{\dagger}(x_6)U(x_1)U^{\dagger}(x_2)
\right]\rangle_Y\nn\\
&&+{\cal K}_{(4;4)}^{(3;5)}\langle
{\rm Tr}\left[U^{\dagger}(x_4)U(z)\right]
{\rm Tr}\left[U^{\dagger}(z)U(x_5)U^{\dagger}(x_6)U(x_1)U^{\dagger}(x_2)U(x_3)
\right]\rangle_Y\nn\\
&&+{\cal K}_{(5;5)}^{(4;6)}\langle
{\rm Tr}\left[U(x_5)U^{\dagger}(z)\right]
{\rm Tr}\left[U(z)U^{\dagger}(x_6)U(x_1)U^{\dagger}(x_2)U(x_3)U^{\dagger}(x_4)
\right]\rangle_Y\nn\\
&&+{\cal K}_{(6;6)}^{(5;1)}\langle
{\rm Tr}\left[U^{\dagger}(x_6)U(z)\right]
{\rm Tr}\left[U^{\dagger}(z)U(x_1)U^{\dagger}(x_2)U(x_3)U^{\dagger}(x_4)U(x_5)
\right]\rangle_Y\nn\\
&&+{\cal K}_{(1;3)}^{(6;4)}\langle
{\rm Tr}\left[U(x_1)U^{\dagger}(x_2)U(x_3)U^{\dagger}(z)\right]
{\rm Tr}\left[U(z)U^{\dagger}(x_4)U(x_5)U^{\dagger}(x_6)\right]\rangle_Y\nn\\
&&+{\cal K}_{(3;5)}^{(2;6)}\langle
{\rm Tr}\left[U(x_3)U^{\dagger}(x_4)U(x_5)U^{\dagger}(z)\right]
{\rm Tr}\left[U(z)U^{\dagger}(x_6)U(x_1)U^{\dagger}(x_2)\right]\rangle_Y\nn\\
&&+{\cal K}_{(5;1)}^{(4;2)}\langle
{\rm Tr}\left[U(x_5)U^{\dagger}(x_6)U(x_1)U^{\dagger}(z)\right]
{\rm Tr}\left[U(z)U^{\dagger}(x_2)U(x_3)U^{\dagger}(x_4)\right]\rangle_Y\nn\\
&&+{\cal K}_{(1;2)}^{(6;3)}\langle
{\rm Tr}\left[U(x_1)U^{\dagger}(x_2)\right]
{\rm Tr}\left[U(x_3)U^{\dagger}(x_4)U(x_5)U^{\dagger}(x_6)\right]\rangle_Y\nn\\
&&+{\cal K}_{(2;3)}^{(1;4)}\langle
{\rm Tr}\left[U^{\dagger}(x_2)U(x_3)\right]
{\rm Tr}\left[U^{\dagger}(x_4)U(x_5)U^{\dagger}(x_6)U(x_1)\right]\rangle_Y\nn\\
&&+{\cal K}_{(3;4)}^{(2;5)}\langle
{\rm Tr}\left[U(x_3)U^{\dagger}(x_4)\right]
{\rm Tr}\left[U(x_5)U^{\dagger}(x_6)U(x_1)U^{\dagger}(x_2)\right]\rangle_Y\nn\\
&&+{\cal K}_{(4;5)}^{(3;6)}\langle
{\rm Tr}\left[U^{\dagger}(x_4)U(x_5)\right]
{\rm Tr}\left[U^{\dagger}(x_6)U(x_1)U^{\dagger}(x_2)U(x_3)\right]\rangle_Y\nn\\
&&+{\cal K}_{(5;6)}^{(4;1)}\langle
{\rm Tr}\left[U(x_5)U^{\dagger}(x_6)\right]
{\rm Tr}\left[U(x_1)U^{\dagger}(x_2)U(x_3)U^{\dagger}(x_4)\right]\rangle_Y\nn\\
&&+{\cal K}_{(6;1)}^{(5;2)}\langle
{\rm Tr}\left[U^{\dagger}(x_6)U(x_1)\right]
{\rm Tr}\left[U^{\dagger}(x_2)U(x_3)U^{\dagger}(x_4)U(x_5)\right]\rangle_Y\nn\\
&&-{\cal P}_{6}
{\rm Tr}\left[U(x_1)U^{\dagger}(x_2)U(x_3)U^{\dagger}(x_4)U(x_5)U^{\dagger}(x_6)
\right]\rangle_Y\nn\\
\end{eqnarray}

 \subsection{Octupole}
By operating functional derivative on the eight wilson line correlator we get
  \begin{eqnarray}
  && \frac{\delta}{\delta \alpha_{u}^a}\frac{\delta}{\delta 
   \alpha_{v}^b}U_{x_1}U_{x_2}^{\dagger}U_{x_3}
  U_{x_4}^{\dagger}U_{x_5}U_{x_6}^{\dagger}U_{x_7}U_{x_8}^{\dagger}= \nn \\
  &+& g^2\delta_{x_1u}\delta_{x_2v} t^a
  U_{x_1}U_{x_2}^{\dagger}
  t^b
  U_{x_3}
  U_{x_4}^{\dagger}U_{x_5}U_{x_6}^{\dagger}U_{x_7}U_{x_8}^{\dagger}
  +  g^2 \delta_{x_1v}\delta_{x_2u} t^b
 U_{x_1}U_{x_2}^{\dagger}
  t^a
  U_{x_3}
  U_{x_4}^{\dagger}U_{x_5}U_{x_6}^{\dagger}U_{x_7}U_{x_8}^{\dagger}
  \nn \\
  &-& g^2\delta_{x_1u}\delta_{x_2v} t^b
  U_{x_1}U_{x_2}^{\dagger}
  t^a
  U_{x_3}
  U_{x_4}^{\dagger}U_{x_5}U_{x_6}^{\dagger}U_{x_7}U_{x_8}^{\dagger}
  - g^2 \delta_{x_1v}\delta_{x_2u} t^a
 U_{x_1}U_{x_2}^{\dagger}
  t^b
  U_{x_3}
  U_{x_4}^{\dagger}U_{x_5}U_{x_6}^{\dagger}U_{x_7}U_{x_8}^{\dagger}
  \nn \\
  &+&   g^2 \delta_{x_1v}\delta_{x_4u} t^b
 U_{x_1}U_{x_2}^{\dagger} U_{x_3}
  U_{x_4}^{\dagger} 
  t^a
 U_{x_5}U_{x_6}^{\dagger}U_{x_7}U_{x_8}^{\dagger}
  +g^2  \delta_{x_1u}\delta_{x_4v} t^a
 U_{x_1}U_{x_2}^{\dagger} U_{x_3}
  U_{x_4}^{\dagger}
  t^b
 U_{x_5}U_{x_6}^{\dagger}U_{x_7}U_{x_8}^{\dagger}
  \nn \\
  &+& g^2\delta_{x_2v}\delta_{x_3u}
   U_{x_1}U_{x_2}^{\dagger}
  t^b t^a
  U_{x_3}
  U_{x_4}^{\dagger}   U_{x_5}U_{x_6}^{\dagger}U_{x_7}U_{x_8}^{\dagger}
 + g^2  \delta_{x_2u}\delta_{x_3v}
  U_{x_1}U_{x_2}^{\dagger}
 t^at^b
  U_{x_3}
  U_{x_4}^{\dagger}   U_{x_5}U_{x_6}^{\dagger}U_{x_7}U_{x_8}^{\dagger}
  \nonumber\\
  &-& g^2 
  \delta_{x_2v}\delta_{x_4u} 
   U_{x_1}U_{x_2}^{\dagger}
  t^b
 U_{x_3}
  U_{x_4}^{\dagger} 
  t^a
  U_{x_5}U_{x_6}^{\dagger}U_{x_7}U_{x_8}^{\dagger}
-
   g^2 
  \delta_{x_2u}\delta_{x_4v} 
  U_{x_1}U_{x_2}^{\dagger}t^a U_{x_3}
  U_{x_4}^{\dagger}  
  t^b
  U_{x_5}U_{x_6}^{\dagger}U_{x_7}U_{x_8}^{\dagger} \nonumber\\
  &+& g^2  \delta_{x_3u}\delta_{x_4v}  U_{x_1}U_{x_2}^{\dagger}t^a U_{x_3}
  U_{x_4}^{\dagger} t^b
  U_{x_5}U_{x_6}^{\dagger}U_{x_7}U_{x_8}^{\dagger}
 + g^2 
  \delta_{x_3v}\delta_{x_4u} U_{x_1}U_{x_2}^{\dagger}t^bU_{x_3}
  U_{x_4}^{\dagger} t^a
 U_{x_5}U_{x_6}^{\dagger}U_{x_7}U_{x_8}^{\dagger} 
 \nn\\
 &-&
   g^2 
  \delta_{x_1v}\delta_{x_5u} t^b U_{x_1}U_{x_2}^{\dagger}U_{x_3}
  U_{x_4}^{\dagger} t^a
 U_{x_5}U_{x_6}^{\dagger}U_{x_7}U_{x_8}^{\dagger} 
- g^2 \delta_{x_1u}\delta_{x_5v} t^a  U_{x_1}U_{x_2}^{\dagger}U_{x_3}
  U_{x_4}^{\dagger} t^b
 U_{x_5}U_{x_6}^{\dagger}U_{x_7}U_{x_8}^{\dagger} 
  \nonumber\\
  &+&
   g^2 
  \delta_{x_1v}\delta_{x_6u} t^b U_{x_1}U_{x_2}^{\dagger}U_{x_3}
  U_{x_4}^{\dagger}
  U_{x_5}U_{x_6}^{\dagger}t^aU_{x_7}U_{x_8}^{\dagger}
+
   g^2 
  \delta_{x_1u}\delta_{x_6v} t^a U_{x_1}U_{x_2}^{\dagger}U_{x_3}
  U_{x_4}^{\dagger} 
   U_{x_5}U_{x_6}^{\dagger}t^bU_{x_7}U_{x_8}^{\dagger}
   \nonumber\\
  &-&
   g^2 
  \delta_{x_1v}\delta_{x_7u} t^bU_{x_1}U_{x_2}^{\dagger}U_{x_3}
  U_{x_4}^{\dagger}
   U_{x_5}U_{x_6}^{\dagger}t^aU_{x_7}U_{x_8}^{\dagger}
-
   g^2 
  \delta_{x_1u}\delta_{x_7v} t^a U_{x_1}U_{x_2}^{\dagger}U_{x_3}
  U_{x_4}^{\dagger} 
  U_{x_5}U_{x_6}^{\dagger}t^bU_{x_7}U_{x_8}^{\dagger}
  \nonumber\\
  &+&
   g^2 
  \delta_{x_1v}\delta_{x_8u} t^bU_{x_1}U_{x_2}^{\dagger}U_{x_3}
  U_{x_4}^{\dagger} 
  U_{x_5}U_{x_6}^{\dagger}U_{x_7}U_{x_8}^{\dagger}t^a
+
   g^2 
  \delta_{x_1u}\delta_{x_8v} t^a U_{x_1}U_{x_2}^{\dagger}U_{x_3}
  U_{x_4}^{\dagger}  
  U_{x_5}U_{x_6}^{\dagger}U_{x_7}U_{x_8}^{\dagger}t^b
  \nonumber\\
  &+&
   g^2 
  \delta_{x_2v}\delta_{x_5u} U_{x_1}U_{x_2}^{\dagger}t^bU_{x_3}
  U_{x_4}^{\dagger} t^a
  U_{x_5}U_{x_6}^{\dagger}U_{x_7}U_{x_8}^{\dagger}
+
   g^2 
  \delta_{x_2u}\delta_{x_5v}  U_{x_1}U_{x_2}^{\dagger}t^aU_{x_3}
  U_{x_4}^{\dagger} t^b
   U_{x_5}U_{x_6}^{\dagger}U_{x_7}U_{x_8}^{\dagger}
  \nonumber\\
  &-&
   g^2 
  \delta_{x_2v}\delta_{x_6u}   U_{x_1}U_{x_2}^{\dagger}t^bU_{x_3}
  U_{x_4}^{\dagger} 
  U_{x_5}U_{x_6}^{\dagger}t^aU_{x_7}U_{x_8}^{\dagger}
-
   g^2 
  \delta_{x_2u}\delta_{x_6v}  U_{x_1}U_{x_2}^{\dagger}t^aU_{x_3}
  U_{x_4}^{\dagger} 
  U_{x_5}U_{x_6}^{\dagger}t^bU_{x_7}U_{x_8}^{\dagger}
   \nonumber\\
   &+&
   g^2 
  \delta_{x_2v}\delta_{x_7u}  U_{x_1}U_{x_2}^{\dagger}t^bU_{x_3}
  U_{x_4}^{\dagger} 
  U_{x_5}U_{x_6}^{\dagger}t^aU_{x_7}U_{x_8}^{\dagger}
 +
   g^2 
  \delta_{x_2u}\delta_{x_7v}   U_{x_1}U_{x_2}^{\dagger}t^aU_{x_3}
  U_{x_4}^{\dagger}
   U_{x_5}U_{x_6}^{\dagger}t^bU_{x_7}U_{x_8}^{\dagger}
\nonumber\\
&-&
   g^2 
  \delta_{x_2v}\delta_{x_8u}   U_{x_1}U_{x_2}^{\dagger}t^bU_{x_3}
  U_{x_4}^{\dagger} 
  U_{x_5}U_{x_6}^{\dagger}U_{x_7}U_{x_8}^{\dagger} t^a 
-
   g^2 
  \delta_{x_2u}\delta_{x_8v}   U_{x_1}U_{x_2}^{\dagger}t^aU_{x_3}
  U_{x_4}^{\dagger} 
  U_{x_5}U_{x_6}^{\dagger}U_{x_7}U_{x_8}^{\dagger} t^b
  \nonumber\\
  &-&
   g^2 
  \delta_{x_3v}\delta_{x_5u}   U_{x_1}U_{x_2}^{\dagger}t^bU_{x_3}
  U_{x_4}^{\dagger} t^a
 U_{x_5}U_{x_6}^{\dagger}U_{x_7}U_{x_8}^{\dagger} 
-
   g^2 
  \delta_{x_3u}\delta_{x_5v}   U_{x_1}U_{x_2}^{\dagger}t^aU_{x_3}
  U_{x_4}^{\dagger} t^b
  U_{x_5}U_{x_6}^{\dagger}U_{x_7}U_{x_8}^{\dagger}
  \nonumber\\
  &+&
   g^2 
  \delta_{x_3v}\delta_{x_6u}   U_{x_1}U_{x_2}^{\dagger}t^bU_{x_3}
  U_{x_4}^{\dagger} 
  U_{x_5}U_{x_6}^{\dagger}t^aU_{x_7}U_{x_8}^{\dagger} 
+
   g^2 
  \delta_{x_3u}\delta_{x_6v}   U_{x_1}U_{x_2}^{\dagger}t^aU_{x_3}
  U_{x_4}^{\dagger} 
  U_{x_5}U_{x_6}^{\dagger}t^bU_{x_7}U_{x_8}^{\dagger}
  \nonumber\\
  &-&
   g^2 
  \delta_{x_3v}\delta_{x_7u}  U_{x_1}U_{x_2}^{\dagger}t^bU_{x_3}
  U_{x_4}^{\dagger} 
  U_{x_5}U_{x_6}^{\dagger}t^aU_{x_7}U_{x_8}^{\dagger}
-
   g^2 
  \delta_{x_3u}\delta_{x_7v}  U_{x_1}U_{x_2}^{\dagger}t^aU_{x_3}
  U_{x_4}^{\dagger} 
 U_{x_5}U_{x_6}^{\dagger}t^bU_{x_7}U_{x_8}^{\dagger}
  \nonumber\\
  &+&
   g^2 
  \delta_{x_3v}\delta_{x_8u}  U_{x_1}U_{x_2}^{\dagger}t^bU_{x_3}
  U_{x_4}^{\dagger}
  U_{x_5}U_{x_6}^{\dagger}U_{x_7}U_{x_8}^{\dagger}t^a
+
   g^2 
  \delta_{x_3u}\delta_{x_8v} U_{x_1}U_{x_2}^{\dagger}t^aU_{x_3}
  U_{x_4}^{\dagger} 
 U_{x_5}U_{x_6}^{\dagger}U_{x_7}U_{x_8}^{\dagger}t^b
  \nonumber\\
  &+&
   g^2 
  \delta_{x_4u}\delta_{x_5v}  U_{x_1}U_{x_2}^{\dagger}U_{x_3}
  U_{x_4}^{\dagger}t^at^b 
  U_{x_5}U_{x_6}^{\dagger}U_{x_7}U_{x_8}^{\dagger}
+
   g^2 
  \delta_{x_4v}\delta_{x_5u}  U_{x_1}U_{x_2}^{\dagger}U_{x_3}
  U_{x_4}^{\dagger}t^bt^a
 U_{x_5}U_{x_6}^{\dagger}U_{x_7}U_{x_8}^{\dagger}
  \nonumber\\
  &-&
   g^2 
  \delta_{x_4u}\delta_{x_6v}  U_{x_1}U_{x_2}^{\dagger}U_{x_3}U_{x_4}^{\dagger}t^a
  U_{x_5}U_{x_6}^{\dagger}t^bU_{x_7}U_{x_8}^{\dagger}
-
   g^2 
  \delta_{x_4v}\delta_{x_6u}  U_{x_1}U_{x_2}^{\dagger}U_{x_3}U_{x_4}^{\dagger}t^b
  U_{x_5}U_{x_6}^{\dagger}t^aU_{x_7}U_{x_8}^{\dagger}
  \nonumber\\
 &+&
   g^2 
  \delta_{x_4u}\delta_{x_7v}  U_{x_1}U_{x_2}^{\dagger}U_{x_3}U_{x_4}^{\dagger}t^a
  U_{x_5}U_{x_6}^{\dagger}t^bU_{x_7}U_{x_8}^{\dagger}
+
   g^2 
  \delta_{x_4v}\delta_{x_7u}  U_{x_1}U_{x_2}^{\dagger}U_{x_3}U_{x_4}^{\dagger}t^b
  U_{x_5}U_{x_6}^{\dagger}t^aU_{x_7}U_{x_8}^{\dagger}
  \nonumber\\
  &-&
   g^2 
  \delta_{x_4u}\delta_{x_8v}  U_{x_1}U_{x_2}^{\dagger}U_{x_3}U_{x_4}^{\dagger}t^a
  U_{x_5}U_{x_6}^{\dagger}U_{x_7}U_{x_8}^{\dagger}t^b
-
   g^2 
  \delta_{x_4v}\delta_{x_8u}  U_{x_1}U_{x_2}^{\dagger}U_{x_3}U_{x_4}^{\dagger}t^b
  U_{x_5}U_{x_6}^{\dagger}U_{x_7}U_{x_8}^{\dagger}t^a  
  \nonumber\\
  &+&
   g^2 
  \delta_{x_5u}\delta_{x_6v}  U_{x_1}U_{x_2}^{\dagger}U_{x_3}U_{x_4}^{\dagger}t^a
  U_{x_5}U_{x_6}^{\dagger}t^bU_{x_7}U_{x_8}^{\dagger}
+
   g^2 
  \delta_{x_5v}\delta_{x_6u}  U_{x_1}U_{x_2}^{\dagger}U_{x_3}U_{x_4}^{\dagger}t^b
  U_{x_5}U_{x_6}^{\dagger}t^aU_{x_7}U_{x_8}^{\dagger}
  \nonumber\\
  &-&
   g^2 
  \delta_{x_5v}\delta_{x_7u}  U_{x_1}U_{x_2}^{\dagger}U_{x_3}U_{x_4}^{\dagger}t^b
  U_{x_5}U_{x_6}^{\dagger}t^aU_{x_7}U_{x_8}^{\dagger}
-
   g^2 
  \delta_{x_5u}\delta_{x_7v}  U_{x_1}U_{x_2}^{\dagger}U_{x_3}U_{x_4}^{\dagger}t^a
  U_{x_5}U_{x_6}^{\dagger}t^bU_{x_7}U_{x_8}^{\dagger}
   \nonumber\\
   &+&
   g^2 
  \delta_{x_5v}\delta_{x_8u}  U_{x_1}U_{x_2}^{\dagger}U_{x_3}U_{x_4}^{\dagger}t^b
  U_{x_5}U_{x_6}^{\dagger}U_{x_7}U_{x_8}^{\dagger}t^a
+
   g^2 
  \delta_{x_5u}\delta_{x_8v}  U_{x_1}U_{x_2}^{\dagger}U_{x_3}U_{x_4}^{\dagger}t^a
  U_{x_5}U_{x_6}^{\dagger}U_{x_7}U_{x_8}^{\dagger}t^b
  \nonumber\\
  &+&
   g^2 
  \delta_{x_6v}\delta_{x_7u}  U_{x_1}U_{x_2}^{\dagger}U_{x_3}U_{x_4}^{\dagger}
  U_{x_5}U_{x_6}^{\dagger}t^bt^aU_{x_7}U_{x_8}^{\dagger}
+
   g^2 
  \delta_{x_6u}\delta_{x_7v}  U_{x_1}U_{x_2}^{\dagger}U_{x_3}U_{x_4}^{\dagger}
  U_{x_5}U_{x_6}^{\dagger}t^at^bU_{x_7}U_{x_8}^{\dagger}
\nonumber\\
 &-&
   g^2 
 \delta_{x_6v}\delta_{x_8u}  U_{x_1}U_{x_2}^{\dagger}U_{x_3}U_{x_4}^{\dagger}
  U_{x_5}U_{x_6}^{\dagger}t^bU_{x_7}U{x_8}^{\dagger}t^a
 -
  g^2 
  \delta_{x_6u}\delta_{x_8v}  U_{x_1}U_{x_2}^{\dagger}U_{x_3}U_{x_4}^{\dagger}
   U_{x_5}U_{x_6}^{\dagger}t^aU_{x_7}U_{x_8}^{\dagger}t^b
   \nonumber\\
   &+&
    g^2 
 \delta_{x_7u}\delta_{x_8v}  U_{x_1}U_{x_2}^{\dagger}U_{x_3}U_{x_4}^{\dagger}
  U_{x_5}U_{x_6}^{\dagger}t^aU_{x_7}U_{x_8}^{\dagger}t^b
+
   g^2 
  \delta_{x_7v}\delta_{x_8u}  U_{x_1}U_{x_2}^{\dagger}U_{x_3}U_{x_4}^{\dagger}
  U_{x_5}U_{x_6}^{\dagger}t^bU_{x_7}U_{x_8}^{\dagger}t^a
  \end{eqnarray}
  Since there are four different terms inside the square bracket in
the definition of JIMWLK Hamiltonian, we are to compute the contributions of these four terms separately. Here we present the final  results of the octupole evolution, 
It can also be seen the coefficient of the non-leading $N _c$
terms vanishes in intermediate steps of the calculations. In the end, we find the sum of all these four contributions leads to the right hand side of Eq. (39) without finite $N_c$ corrections. Finally for octupole ($n=4$), we get, 
\begin{eqnarray}
 &&\frac{\partial}{\partial Y} {\rm Tr}\left[U(x_{1})U^{\dagger}(x_{2})
U(x_{3})U^{\dagger}(x_{4}) U(x_{5})U^{\dagger}(x_{6})U(x_7)U(x_8)^{\dagger}\right] \nn\\
&=&\frac{\bar \alpha_{s}}{4\pi}\int_{z}{\cal K}_{(1;1)}^{(8;2)}\langle
{\rm Tr}\left[U(x_1)U^{\dagger}(z)\right]
{\rm Tr}\left[U(z)U^{\dagger}(x_2)U(x_3)U^{\dagger}(x_4)U(x_5)U^{\dagger}(x_6)U(x_7)U^{\dagger}(x_8)\right]\rangle_Y\nn\\
&&+{\cal K}_{(2;2)}^{(1;3)}\langle
{\rm Tr}\left[U^{\dagger}(x_2)U(z)\right]
{\rm Tr}\left[U^{\dagger}(z)U(x_3)U^{\dagger}(x_4)U(x_5)U^{\dagger}(6)U(x_7)U^{\dagger}(x_8)U(x_1)\right]\rangle_Y\nn\\
&&+{\cal K}_{(3;3)}^{(2;4)}\langle
{\rm Tr}\left[U(x_3)U^{\dagger}(z)\right]
{\rm Tr}\left[U(z)U^{\dagger}(x_4)U(x_5)U^{\dagger}(x_6)U(x_7)U^{\dagger}(x_8)U(x_1)U^{\dagger}(x_2)\right]\rangle_Y\nn\\
&&+{\cal K}_{(4;4)}^{(3;5)}\langle
{\rm Tr}\left[U^{\dagger}(x_4)U(z)\right]
{\rm Tr}\left[U^{\dagger}(z)U(x_5)U^{\dagger}(x_6)U(x_7)U^{\dagger}(8)U(x_1)U^{\dagger}(x_2)U(x_3)\right]\rangle_Y\nn\\
&&+{\cal K}_{(5;5)}^{(4;6)}\langle
{\rm Tr}\left[U(x_5)U^{\dagger}(z)\right]
{\rm Tr}\left[U(z)U^{\dagger}(x_6)U(x_7)U^{\dagger}(x_8)U(x_1)U^{\dagger}(x_2)U(x_3)U^{\dagger}(x_4)\right]\rangle_Y\nn\\
&&+{\cal K}_{(6;6)}^{(5;7)}\langle
{\rm Tr}\left[U^{\dagger}(x_6)U(z)\right]
{\rm Tr}\left[U^{\dagger}(z)U(x_7)U^{\dagger}(x_8)U(x_1)U^{\dagger}(2)U(x_3)U^{\dagger}(x_4)U(x_5)\right]\rangle_Y\nn\\
&&+{\cal K}_{(7;7)}^{(6;8)}\langle
{\rm Tr}\left[U(x_7)U^{\dagger}(z)\right]
{\rm Tr}\left[U(z)U^{\dagger}(x_8)U(x_1)U^{\dagger}(x_2)U(x_3)U^{\dagger}(x_4)U(x_5)U^{\dagger}(x_6)\right]\rangle_Y\nn\\
&&+{\cal K}_{(8;8)}^{(7;1)}\langle
{\rm Tr}\left[U^{\dagger}(x_8)U(z)\right]
{\rm Tr}\left[U^{\dagger}(z)U(x_1)U^{\dagger}(x_2)U(x_3)U^{\dagger}(x_4)U(x_5)U^{\dagger}(x_6)U(x_7)\right]\rangle_Y\nn\\
&&+{\cal K}_{(1;5)}^{(8;6)}\langle{\rm Tr}\left[U(x_1)U^{\dagger}(x_2)U(x_3)U^{\dagger}(x_4)U(x_5)U^{\dagger}(z)\right]
{\rm Tr}\left[U(z)U^{\dagger}(x_6)U(x_7)U^{\dagger}(x_8)\right]
 \rangle_Y\nn\\
 &&+{\cal K}_{(2;6)}^{(7;1)}\langle{\rm Tr}\left[U^{\dagger}(x_2)U(x_3)U^{\dagger}(x_4)U(x_5)U^{\dagger}(x_6)U(z)\right]
{\rm Tr}\left[U^{\dagger}(z)U(x_7)U^{\dagger}(x_8)U(x_1)\right]
 \rangle_Y\nn\\
 &&+{\cal K}_{(3;7)}^{(2;8)}\langle{\rm Tr}\left[U(x_3)U^{\dagger}(x_4)U(x_5)U^{\dagger}(x_6)U(x_7)U^{\dagger}(z)\right]
{\rm Tr}\left[U(z)U^{\dagger}(x_8)U(x_1)U^{\dagger}(x_2)\right]
 \rangle_Y\nn\\
&&+{\cal K}_{(4;8)}^{(3;1)}\langle{\rm Tr}\left[U^{\dagger}(x_4)U(x_5)U^{\dagger}(x_6)U(x_7)U^{\dagger}(x_8)U(z)\right]
{\rm Tr}\left[U^{\dagger}(z)U(x_1)U^{\dagger}(x_2)U(x_3)\right]\rangle_Y\nn\\
&&+{\cal K}_{(5;1)}^{(4;2)}\langle{\rm Tr}\left[U(x_5)U^{\dagger}(x_6)U(x_7)U^{\dagger}(x_8)U(x_1)U^{\dagger}(z)\right]
{\rm Tr}\left[U(z)U^{\dagger}(x_2)U(x_3)U^{\dagger}(x_4)\right]
 \rangle_Y\nn\\
 &&+{\cal K}_{(6;2)}^{(5;3)}\langle{\rm Tr}\left[U^{\dagger}(x_6)U(x_7)U^{\dagger}(x_8)U(x_1)U^{\dagger}(x_2)U(z)\right]
{\rm Tr}\left[U^{\dagger}(z)U(x_3)U^{\dagger}(x_4)U(x_5)\right]
 \rangle_Y\nn\\
 &&+{\cal K}_{(7;3)}^{(6;4)}\langle{\rm Tr}\left[U(x_7)U^{\dagger}(x_8)U(x_1)U^{\dagger}(x_2)U(x_3)U^{\dagger}(z)\right]
{\rm Tr}\left[U(z)U^{\dagger}(x_4)U(x_5)U^{\dagger}(x_6)\right]
 \rangle_Y\nn\\
 &&+{\cal K}_{(8;4)}^{(7;5)}\langle{\rm Tr}\left[U^{\dagger}(x_8)U(x_1)U^{\dagger}(x_2)U(x_3)U^{\dagger}(x_4)U(z)\right]
{\rm Tr}\left[U^{\dagger}(z)U(x_5)U^{\dagger}(x_6)U(x_7)\right]
 \rangle_Y\nn\\
 &&+{\cal K}_{(1;4)}^{(8;5)}\langle{\rm Tr}\left[U(x_1)U^{\dagger}(x_2)U(x_3)U^{\dagger}(x_4)\right]
 {\rm Tr} \left[U(x_5)U^{\dagger}(x_6)U(x_7)U^{\dagger}(x_8)\right] \rangle_Y\nn\\
 &&+{\cal K}_{(2;5)}^{(1;6)}\langle{\rm Tr}\left[U^{\dagger}(x_2)U(x_3)U^{\dagger}(x_4)U(x_5)\right]
{\rm Tr}\left[U^{\dagger}(x_6)U(x_7)U^{\dagger}(x_8)U(x_1)\right]
  \rangle_Y\nn\\
 &&+{\cal K}_{(3;6)}^{(2;7)}\langle{\rm Tr}\left[U(x_3)U^{\dagger}(x_4)U(x_5)U^{\dagger}(x_6)\right]
 {\rm Tr}\left[U(x_7)U^{\dagger}(x_8)U(x_1)U^{\dagger}(x_2)\right]\rangle_Y\nn\\ 
 &&+{\cal K}_{(4;7)}^{(3;8)}\langle{\rm Tr}\left[U^{\dagger}(x_4)U(x_5)U^{\dagger}(x_6)U(x_7)\right]
{\rm Tr}\left[U^{\dagger}(x_8)U(x_1)U^{\dagger}(x_2)U(x_3)\right]
 \rangle_Y\nn\\
 &&+{\cal K}_{(1;2)}^{(8;3)}\langle
{\rm Tr}\left[U(x_1)U^{\dagger}(x_2)\right]
{\rm Tr}\left[U(x_3)U^{\dagger}(x_4)U(x_5)U^{\dagger}(x_6)U(x_7)U^{\dagger}(x_8)\right]\rangle_Y\nn\\
&&+{\cal K}_{(2;3)}^{(1;4)}\langle
{\rm Tr}\left[U^{\dagger}(x_2)U(x_3)\right]
{\rm Tr}\left[U^{\dagger}(x_4)U(x_5)U^{\dagger}(x_6)U(x_7)U^{\dagger}(x_8)U(x_1)\right]\rangle_Y\nn\\
&&+{\cal K}_{(3;4)}^{(2;5)}\langle
{\rm Tr}\left[U(x_3)U^{\dagger}(x_4)\right]
{\rm Tr}\left[U(x_5)U^{\dagger}(x_6)U(x_7)U^{\dagger}(x_8)U(x_1)U^{\dagger}(x_2)\right]\rangle_Y\nn\\
&&+{\cal K}_{(4;5)}^{(3;6)}\langle
{\rm Tr}\left[U^{\dagger}(x_4)U(x_5)\right]
{\rm Tr}\left[U^{\dagger}(x_6)U(x_7)U^{\dagger}(x_8)U(x_1)U^{\dagger}(x_2)U(x_3)\right]\rangle_Y\nn\\
&&+{\cal K}_{(5;6)}^{(4;7)}\langle
{\rm Tr}\left[U(x_5)U^{\dagger}(x_6)\right]
 {\rm Tr}\left[U(x_7)U^{\dagger}(x_8)U(x_1)U^{\dagger}(x_2)U(x_3)U^{\dagger}(x_4)\right]\rangle_Y\nn\\
&&+{\cal K}_{(6;7)}^{(5;8)}\langle
{\rm Tr}\left[U(x_6)U^{\dagger}(x_7)\right]
{\rm Tr}\left[U^{\dagger}(x_8)U(x_1)U^{\dagger}(x_2)U(x_3)U^{\dagger}(x_4)U(x_5)\right]\rangle_Y\nn\\
&&+{\cal K}_{(7;8)}^{(6;1)}\langle
{\rm Tr}\left[U^{\dagger}(x_7)U(x_8)\right]
{\rm Tr}\left[U(x_1)U^{\dagger}(x_2)U(x_3)U^{\dagger}(x_4)U(x_5)U^{\dagger}(x_6)\right]\rangle_Y \nn\\
&&+{\cal K}_{(8;1)}^{(7;2)}\langle
{\rm Tr}\left[U(x_8)U^{\dagger}(x_1)\right]
{\rm Tr}\left[U^{\dagger}(x_2)U(x_3)U^{\dagger}(x_4)U(x_5)U^{\dagger}(x_6)U(x_7)\right]\rangle_Y\nn\\
 &&-{\cal P}_{8} \langle
{\rm Tr}\left[U(x_1)U^{\dagger}(x_2)U(x_3)U^{\dagger}(x_4)U(x_5)U^{\dagger}(x_6)U(x_7)U^{\dagger}(x_8)\right]\rangle_Y 
\label{eq2}
 \end{eqnarray}

\subsection{Decapole}
For decapole $n=5$,
\begin{eqnarray}
 &&\frac{\partial}{\partial Y} {\rm Tr}\left[U(x_{1})U^{\dagger}(x_{2})
U(x_{3})U^{\dagger}(x_{4}) U(x_{5})U^{\dagger}(x_{6}U(x_{7})U^{\dagger}(x_{8})
U(x_{7})U^{\dagger}(x_{8}) U(x_{9})U^{\dagger}(x_{10})\right] \nn\\
&=&\frac{\bar \alpha_{s}}{4\pi}\int_{z}{\cal K}_{(1;1)}^{(10;2)}\langle
{\rm Tr}\left[U(x_1)U^{\dagger}(z)\right]
{\rm Tr}\left[U(z)U^{\dagger}(x_2)U(x_3)U^{\dagger}(x_4)U(x_5)
U^{\dagger}(x_6)U(x_7)U^{\dagger}(x_8)U(x_{9})U^{\dagger}(x_{10})\right]\rangle_Y\nn\\
&&+{\cal K}_{(2;2)}^{(1;3)}\langle
{\rm Tr}\left[U^{\dagger}(x_2)U(z)\right]
{\rm Tr}\left[U^{\dagger}(z)U(x_3)U^{\dagger}(x_4)U(x_5)U^{\dagger}(x_6)U(x_7)
U^{\dagger}(x_8)U(x_9)U^{\dagger}(x_{10})U(x_1)\right]\rangle_Y\nn\\
&&+{\cal K}_{(3;3)}^{(2;4)}\langle
{\rm Tr}\left[U(x_3)U^{\dagger}(z)\right]
{\rm Tr}\left[U(z)U^{\dagger}(x_4)U(x_5)U^{\dagger}(x_6)U(x_7)U^{\dagger}(x_8)U(x_9)
U^{\dagger}(x_{10})U(x_1)U^{\dagger}(x_2)\right]\rangle_Y\nn\\
&&+{\cal K}_{(4;4)}^{(3;5)}\langle
{\rm Tr}\left[U^{\dagger}(x_4)U(z)\right]
{\rm Tr}\left[U^{\dagger}(z)U(x_5)U^{\dagger}(x_6)U(x_7)U^{\dagger}(x_8)U(x_9)U^{\dagger}(x_{10})
U(x_1)U^{\dagger}(x_2)U(x_3)\right]\rangle_Y\nn\\
&&+{\cal K}_{(5;5)}^{(4;6)}\langle
{\rm Tr}\left[U(x_5)U^{\dagger}(z)\right]
{\rm Tr}\left[U(z)U^{\dagger}(x_6)U(x_7)U^{\dagger}(x_8)U(x_9)U^{\dagger}(x_{10})U(x_1)
U^{\dagger}(x_2)U(x_3)U^{\dagger}(x_4)\right]\rangle_Y\nn\\
&&+{\cal K}_{(6;6)}^{(5;7)}\langle
{\rm Tr}\left[U^{\dagger}(x_6)U(z)\right]
{\rm Tr}\left[U^{\dagger}(z)U(x_7)U^{\dagger}(x_8)U(x_9)U^{\dagger}(x_{10})U(x_1)
U^{\dagger}(x_2)U(x_3)U^{\dagger}(x_4)U(x_5)\right]\rangle_Y\nn\\
&&+{\cal K}_{(7;7)}^{(6;8)}\langle
{\rm Tr}\left[U(x_7)U^{\dagger}(z)\right]
{\rm Tr}\left[U(z)U^{\dagger}(x_8)U(x_9)U^{\dagger}(x_{10})U(x_1)U^{\dagger}
(x_2)U(x_3)U^{\dagger}(x_4)U(x_5)U^{\dagger}(x_6)\right]\rangle_Y\nn\\
&&+{\cal K}_{(8;8)}^{(7;9)}\langle
{\rm Tr}\left[U^{\dagger}(x_8)U(z)\right]
{\rm Tr}\left[U^{\dagger}(z)U(x_9)U^{\dagger}(x_{10})U(x_1)U^{\dagger}(x_2)U(x_3)
U^{\dagger}(x_4)U(x_5)U^{\dagger}(x_6)U(x_7)\right]\rangle_Y\nn\\
&&+{\cal K}_{(9;9)}^{(8;10)}\langle
{\rm Tr}\left[U(x_9)U^{\dagger}(z)\right]
{\rm Tr}\left[U(z)U^{\dagger}(x_{10})U(x_1)U^{\dagger}(x_2)U(x_3)
U^{\dagger}(x_4)U(x_5)U^{\dagger}(x_6)U(x_7)U^{\dagger}(x_8)\right]\rangle_Y\nn\\
&&+{\cal K}_{(10;10)}^{(9;1)}\langle
{\rm Tr}\left[U^{\dagger}(x_{10})U(z)\right]
{\rm Tr}\left[U^{\dagger}(z)U(x_1)U^{\dagger}(x_2)U(x_3)
U^{\dagger}(x_4)U(x_5)U^{\dagger}(x_6)U(x_7)U^{\dagger}(x_8)U(x_9)\right]\rangle_Y\nn\\
&&+{\cal K}_{(1;5)}^{(10;6)}\langle
{\rm Tr}\left[U(x_1)U^{\dagger}(x_2)U(x_3)U^{\dagger}(x_4)U(x_5)U^{\dagger}(z)\right]
{\rm Tr}\left[U(z)U^{\dagger}(x_6)U(x_7)U^{\dagger}(x_8)U(x_9)
U^{\dagger}(x_{10})\right]\rangle_Y\nn\\
&&+{\cal K}_{(2;6)}^{(1;7)}\langle
{\rm Tr}\left[U^{\dagger}(x_2)U(x_3)U^{\dagger}(x_4)U(x_5)U^{\dagger}(x_6)U(z)\right]
{\rm Tr}\left[U^{\dagger}(z)U(x_7)U^{\dagger}(x_8)U(x_9)
U^{\dagger}(x_{10})U(x_1)\right]\rangle_Y\nn\\
&&+{\cal K}_{(3;7)}^{(2;8)}\langle
{\rm Tr}\left[U(x_3)U^{\dagger}(x_4)U(x_5)U^{\dagger}(x_6)U(x_7)U^{\dagger}(z)\right]
{\rm Tr}\left[U(z)U^{\dagger}(x_8)U(x_9)U^{\dagger}(x_{10})U(x_1)U^{\dagger}(x_2)\right]
\rangle_Y\nn\\
&&+{\cal K}_{(4;8)}^{(3;9)}\langle
{\rm Tr}\left[U^{\dagger}(x_4)U(x_5)U^{\dagger}(x_6)U(x_7)U^{\dagger}(x_8)U(z)\right]
{\rm Tr}\left[U^{\dagger}(z)U(x_9)U^{\dagger}(x_{10})U(x_1)U^{\dagger}(x_2)U(x_3)
\right]\rangle_Y\nn\\
&&+{\cal K}_{(5;9)}^{(4;10)}\langle
{\rm Tr}\left[U(x_5)U^{\dagger}(x_6)U(x_7)U^{\dagger}(x_8)U(x_9)U^{\dagger}(z)\right]
{\rm Tr}\left[U(z)U^{\dagger}(x_{10})U(x_1)U^{\dagger}(x_2)U(x_3)U^{\dagger}(x_4)\right]
\rangle_Y\nn\\
&&+{\cal K}_{(1;7)}^{(10;8)}\langle
{\rm Tr}\left[U(x_1)U^{\dagger}(x_2)U(x_3)U^{\dagger}(x_4)U(x_5)
U^{\dagger}(x_6)U(x_7)U^{\dagger}(z)\right]
{\rm Tr}\left[U(z)U^{\dagger}(x_8)U(x_9)
U^{\dagger}(x_{10})\right]\rangle_Y\nn\\
&&+{\cal K}_{(2;8)}^{(1;9)}\langle
{\rm Tr}\left[U^{\dagger}(x_2)U(x_3)U^{\dagger}(x_4)U(x_5)U^{\dagger}(x_6)U(x_7)
U^{\dagger}(x_8)U(z)\right]
{\rm Tr}\left[U^{\dagger}(z)U(x_9)U^{\dagger}(x_{10})U(x_1)
\right]\rangle_Y\nn\\
&&+{\cal K}_{(3;9)}^{(2;10)}\langle
{\rm Tr}\left[U(x_3)U^{\dagger}(x_4)U(x_5)U^{\dagger}(x_6)U(x_7)
U^{\dagger}(x_8)U(x_9)U^{\dagger}(z)\right]
{\rm Tr}\left[U(z)U^{\dagger}(x_{10})U(x_1)
U^{\dagger}(x_2)\right]\rangle_Y\nn\\
&&+{\cal K}_{(4;10)}^{(3;1)}\langle
{\rm Tr}\left[U^{\dagger}(x_4)U(x_5)U^{\dagger}(x_6)U(x_7)U^{\dagger}(x_8)U(x_9)
U^{\dagger}(x_{10})U(z)\right]
{\rm Tr}\left[U^{\dagger}(z)U(x_1)U^{\dagger}(x_2)U(x_3)
\right]\rangle_Y\nn\\
&&+{\cal K}_{(5;1)}^{(4;2)}\langle
{\rm Tr}\left[U(x_5)U^{\dagger}(x_6)U(x_7)U^{\dagger}(x_8)U(x_9)
U^{\dagger}(x_{10})U(x_1)U^{\dagger}(z)\right]
{\rm Tr}\left[U(z)U^{\dagger}(x_2)U(x_3)
U^{\dagger}(x_4)\right]\rangle_Y\nn\\
&&+{\cal K}_{(6;2)}^{(5;3)}\langle
{\rm Tr}\left[U^{\dagger}(x_6)U(x_7)U^{\dagger}(x_8)U(x_9)U^{\dagger}(x_{10})U(x_1)
U^{\dagger}(x_2)U(z)\right]
{\rm Tr}\left[U^{\dagger}(z)U(x_3)U^{\dagger}(x_4)U(x_5)
\right]\rangle_Y\nn\\
&&+{\cal K}_{(7;3)}^{(6;4)}\langle
{\rm Tr}\left[U(x_7)U^{\dagger}(x_8)U(x_9)U^{\dagger}(x_{10})U(x_1)
U^{\dagger}(x_2)U(x_3)U^{\dagger}(z)\right]
{\rm Tr}\left[U(z)U^{\dagger}(x_4)U(x_5)
U^{\dagger}(x_6)\right]\rangle_Y\nn\\
&&+{\cal K}_{(8;4)}^{(7;5)}\langle
{\rm Tr}\left[U^{\dagger}(x_8)U(x_9)U^{\dagger}(x_{10})U(x_1)U^{\dagger}(x_2)U(x_3)
U^{\dagger}(x_4)U(z)\right]
{\rm Tr}\left[U^{\dagger}(z)U(x_5)U^{\dagger}(x_6)U(x_7)
\right]\rangle_Y\nn\\
&&+{\cal K}_{(9;5)}^{(8;6)}\langle
{\rm Tr}\left[U(x_9)U^{\dagger}(x_{10})U(x_1)U^{\dagger}(x_2)U(x_3)
U^{\dagger}(x_4)U(x_5)U^{\dagger}(z)\right]
{\rm Tr}\left[U(z)U^{\dagger}(x_6)U(x_7)
U^{\dagger}(x_8)\right]\rangle_Y\nn\\
&&+{\cal K}_{(10;6)}^{(9;7)}\langle
{\rm Tr}\left[U^{\dagger}(x_{10})U(x_1)U^{\dagger}(x_2)U(x_3)U^{\dagger}(x_4)U(x_5)
U^{\dagger}(x_6)U(z)\right]
{\rm Tr}\left[U^{\dagger}(z)U(x_7)U^{\dagger}(x_8)U(x_9)
\right]\rangle_Y\nn\\
&&+{\cal K}_{(1;2)}^{(10;3)}\langle
{\rm Tr}\left[U(x_1)U^{\dagger}(x_2)\right]
{\rm Tr}\left[U(x_3)U^{\dagger}(x_4)U(x_5)
U^{\dagger}(x_6)U(x_7)U^{\dagger}(x_8)U(x_{9})U^{\dagger}(x_{10})\right]\rangle_Y\nn\\
&&+{\cal K}_{(2;3)}^{(1;4)}\langle
{\rm Tr}\left[U^{\dagger}(x_2)U(x_3)\right]
{\rm Tr}\left[U^{\dagger}(x_4)U(x_5)U^{\dagger}(x_6)U(x_7)
U^{\dagger}(x_8)U(x_9)U^{\dagger}(x_{10})U(x_1)\right]\rangle_Y\nn\\
&&+{\cal K}_{(3;4)}^{(2;5)}\langle
{\rm Tr}\left[U(x_3)U^{\dagger}(x_4)\right]
{\rm Tr}\left[U(x_5)U^{\dagger}(x_6)U(x_7)U^{\dagger}(x_8)U(x_9)
U^{\dagger}(x_{10})U(x_1)U^{\dagger}(x_2)\right]\rangle_Y\nn\\
&&+{\cal K}_{(4;5)}^{(3;6)}\langle
{\rm Tr}\left[U^{\dagger}(x_4)U(x_5)\right]
{\rm Tr}\left[U^{\dagger}(x_6)U(x_7)U^{\dagger}(x_8)U(x_9)U^{\dagger}(x_{10})
U(x_1)U^{\dagger}(x_2)U(x_3)\right]\rangle_Y\nn\\
&&+{\cal K}_{(5;6)}^{(4;7)}\langle
{\rm Tr}\left[U(x_5)U^{\dagger}(x_6)\right]
{\rm Tr}\left[U(x_7)U^{\dagger}(x_8)U(x_9)U^{\dagger}(x_{10})U(x_1)
U^{\dagger}(x_2)U(x_3)U^{\dagger}(x_4)\right]\rangle_Y\nn\\
&&+{\cal K}_{(6;7)}^{(5;8)}\langle
{\rm Tr}\left[U^{\dagger}(x_6)U(x_7)\right]
{\rm Tr}\left[U^{\dagger}(x_8)U(x_9)U^{\dagger}(x_{10})U(x_1)
U^{\dagger}(x_2)U(x_3)U^{\dagger}(x_4)U(x_5)\right]\rangle_Y\nn\\
&&+{\cal K}_{(7;8)}^{(6;9)}\langle
{\rm Tr}\left[U(x_7)U^{\dagger}(x_8)\right]
{\rm Tr}\left[U(x_9)U^{\dagger}(x_{10})U(x_1)U^{\dagger}
(x_2)U(x_3)U^{\dagger}(x_4)U(x_5)U^{\dagger}(x_6)\right]\rangle_Y\nn\\
&&+{\cal K}_{(8;9)}^{(7;10)}\langle
{\rm Tr}\left[U^{\dagger}(x_8)U(x_9)\right]
{\rm Tr}\left[U^{\dagger}(x_{10})U(x_1)U^{\dagger}(x_2)U(x_3)
U^{\dagger}(x_4)U(x_5)U^{\dagger}(x_6)U(x_7)\right]\rangle_Y\nn\\
&&+{\cal K}_{(9;10)}^{(8;1)}\langle
{\rm Tr}\left[U(x_9)U^{\dagger}(x_{10})\right]
{\rm Tr}\left[U(x_1)U^{\dagger}(x_2)U(x_3)
U^{\dagger}(x_4)U(x_5)U^{\dagger}(x_6)U(x_7)U^{\dagger}(x_8)\right]\rangle_Y\nn\\
&&+{\cal K}_{(10;1)}^{(9;2)}\langle
{\rm Tr}\left[U^{\dagger}(x_{10})U(x_1)\right]
{\rm Tr}\left[U^{\dagger}(x_2)U(x_3)
U^{\dagger}(x_4)U(x_5)U^{\dagger}(x_6)U(x_7)
U^{\dagger}(x_8)U(x_9)\right]\rangle_Y\nn\\
&&+{\cal K}_{(1;6)}^{10;7)}\langle
{\rm Tr}\left[U(x_1)U^{\dagger}(x_2)U(x_3)
U^{\dagger}(x_4)U(x_5)U^{\dagger}(x_6)\right]
{\rm Tr}\left[U(x_7)U^{\dagger}(x_8)U(x_9)U^{\dagger}(x_{10})\right]\rangle_Y\nn\\
&&+{\cal K}_{(2;7)}^{(1;8)}\langle
{\rm Tr}\left[U^{\dagger}(x_2)U(x_3)
U^{\dagger}(x_4)U(x_5)U^{\dagger}(x_6)U(x_7)\right]
{\rm Tr}\left[U^{\dagger}(x_8)U(x_9)U^{\dagger}(x_{10})U(x_1)
\right]\rangle_Y\nn\\
&&+{\cal K}_{(3;8)}^{(2;9)}\langle
{\rm Tr}\left[U(x_3)U^{\dagger}(x_4)U(x_5)
U^{\dagger}(x_6)U(x_7)U^{\dagger}(x_8)\right]
{\rm Tr}\left[U(x_9)U^{\dagger}(x_{10})U(x_1)U^{\dagger}(x_2)\right]\rangle_Y\nn\\
&&+{\cal K}_{(4;9)}^{(3;10)}\langle
{\rm Tr}\left[U^{\dagger}(x_4)U(x_5)
U^{\dagger}(x_6)U(x_7)U^{\dagger}(x_8)U(x_9)\right]
{\rm Tr}\left[U^{\dagger}(x_{10})U(x_1)U^{\dagger}(x_2)U(x_3)
\right]\rangle_Y\nn\\
&&+{\cal K}_{(5;10)}^{(4;1)}\langle
{\rm Tr}\left[U(x_5)U^{\dagger}(x_6)U(x_7)
U^{\dagger}(x_8)U(x_9)U^{\dagger}(x_{10})\right]
{\rm Tr}\left[U(x_1)U^{\dagger}(x_2)U(x_3)U^{\dagger}(x_4)\right]\rangle_Y\nn\\
&&+{\cal K}_{(6;1)}^{(5;2)}\langle
{\rm Tr}\left[U^{\dagger}(x_6)U(x_7)
U^{\dagger}(x_8)U(x_9)U^{\dagger}(x{10})U(x_1)\right]
{\rm Tr}\left[U^{\dagger}(x_2)U(x_3)U^{\dagger}(x_4)U(x_5)
\right]\rangle_Y\nn\\
&&+{\cal K}_{(7;2)}^{(6;3)}\langle
{\rm Tr}\left[U(x_7)U^{\dagger}(x_8)U(x_9)
U^{\dagger}(x_{10})U(x_1)U^{\dagger}(x_2)\right]
{\rm Tr}\left[U(x_3)U^{\dagger}(x_4)U(x_5)U^{\dagger}(x_6)\right]\rangle_Y\nn\\
&&+{\cal K}_{(8;3)}^{(7;4)}\langle
{\rm Tr}\left[U^{\dagger}(x_8)U(x_9)
U^{\dagger}(x_{10})U(x_1)U^{\dagger}(x_2)U(x_3)\right]
{\rm Tr}\left[U^{\dagger}(x_4)U(x_5)U^{\dagger}(x_6)U(x_7)
\right]\rangle_Y\nn\\
&&+{\cal K}_{(9;4)}^{(8;5)}\langle
{\rm Tr}\left[U(x_9)U^{\dagger}(x_{10})U(x_1)
U^{\dagger}(x_2)U(x_3)U^{\dagger}(x_4)\right]
{\rm Tr}\left[U(x_5)U^{\dagger}(x_6)U(x_7)U^{\dagger}(x_8)\right]\rangle_Y\nn\\
&&+{\cal K}_{(10;5)}^{(9;6)}\langle
{\rm Tr}\left[U^{\dagger}(x_{10})U(x_1)
U^{\dagger}(x_2)U(x_3)U^{\dagger}(x_4)U(x_5)\right]
{\rm Tr}\left[U^{\dagger}(x_6)U(x_7)U^{\dagger}(x_8)U(x_9)
\right]\rangle_Y\nn\\
&&-{\cal P}_{10}\langle
{\rm Tr}\left[U(x_1)U^{\dagger}(x_2)U(x_3)U^{\dagger}(x_4)U(x_5)U^{\dagger}(x_6)U(x_7)
U^{\dagger}(x_8)U(x_9)
U^{\dagger}(x_{10})\right]\rangle_Y\nn\
\end{eqnarray}

\end{document}